\documentclass[12pt]{article}
\usepackage{a4wide}
\usepackage{cite}
\usepackage{latexsym}
\usepackage{amsfonts}
\usepackage{amssymb,amsbsy}
\usepackage{bbm}
\usepackage{slashed}
\usepackage{subeqn}
\usepackage{nonfloat}
\usepackage{morefloats}
\usepackage[dvips]{graphicx}
\renewcommand{\theequation}{\thesection.\arabic{equation}}
\def\vec#1{\boldsymbol{#1}}

\def\ppLL{\ensuremath{\bar{\mathrm{p}}\mathrm{p}\to\overline{\Lambda}\Lambda}}

\makeatletter
\newlength{\earraycolsep}
\def\eqnarray{\stepcounter{equation}\let\@currentlabel%
\theequation
\global\@eqnswtrue\m@th
\global\@eqcnt\z@\tabskip\@centering\let\\\@eqncr
$$\halign to\displaywidth\bgroup\@eqnsel\hskip\@centering
$\displaystyle\tabskip\z@{##}$&\global\@eqcnt\@ne
\hskip 2\earraycolsep \hfil$\displaystyle{##}$\hfil
&\global\@eqcnt\tw@ \hskip 2\earraycolsep
$\displaystyle\tabskip\z@{##}$\hfil
\tabskip\@centering&\llap{##}\tabskip\z@\cr}
\makeatother
\catcode`\@=11
\def\eqalign#1{\null\,\vcenter{\openup\jot\m@th
\ialign{\strut\hfil$\displaystyle{##}$&$\displaystyle{{}##}$\hfil
     \crcr#1\crcr}}\,}
\catcode`\@=12
\catcode`\@=11
\def\Eqalign#1{\null\,\vcenter{\openup\jot\m@th
\ialign{\strut\hfil$\displaystyle{##}$&$\displaystyle{{}##}$\hfil
&&\qquad\strut\hfil$\displaystyle{##}$&$\displaystyle{{}##}$\hfil
     \crcr#1\crcr}}\,}
\catcode`\@=12

\def\acc{\check{\rho}}

\def\Rpt{{\tilde R}}
\def\R{R}

\def\acc{\check\rho}
\def\une{\mathbbm{1}}

\def\R{R}
\def\D{{\cal D}}
\def\<{\langle}
\def\>{\rangle}
\let\tr=\Tr

\def\acc{\check{\rho}}

\def\Sv{{\vec{S}}}

\def\xu{\hat{\vec{x}}}
\def\yu{\hat{\vec{y}}}
\def\zu{\hat{\vec{z}}}
\def\acc{\check{\rho}}

\def\Xrond{{\rm X}\raise1.8ex\hbox{\kern-0.6em\hbox{$\circ$}}\ }
\def\krond{{\bf k}\raise1.8ex\hbox{\kern-0.6em\hbox{$\circ$}}\ }

\def\Ocal{{\mathcal{O}}}
\def\be{\begin{equation}}

\def\ee{\end{equation}}

\def\barr{\begin{eqnarray}}
\def\earr{\end{eqnarray}}

\def\ni{\noindent}

\def\sq{\raise -2pt\hbox{\Large $\Box$}}
\def\dis{\scalebox{.9}{$\bigcirc$}}
\def\disp{\raise -1pt\hbox{\large$\otimes$}}
\def\trv{ \raise -2pt \hbox{\large$\nabla$}}
\def\trh{\raise -2pt \hbox{\rotatebox{90}{\hbox{\large$\Delta$}}}}

\begin{document}
\begin{flushright}
 LPSC 07-201\\
 IPNL xxxxxx
 \end{flushright}
\begin{center}
{\bfseries 
POSITIVITY DOMAINS FOR PAIRS OR TRIPLES OF SPIN OBSERVABLES
\footnote{Talk given by Xavier Artru at ``DSPIN-07'', XII   Workshop on High-Energy Spin Physics,  Dubna, Sept. 3--7, 2007, to appear in the Proceedings}
}
\vskip 5mm
\underline{X. Artru}$^{1 \dag}$, J.M. Richard$^{2}$ and J. Soffer$^{3}$
\vskip 5mm
{\small
(1) {\it Institut de Physique Nucl\'eaire de Lyon, Universit\'e de Lyon,}\\
{\it CNRS- IN2P3 and Universit\'e Lyon 1, 69622 Villeurbanne, France}\\
(2) {\it Laboratoire de Physique Subatomique et Cosmologie,}\\
{\it CNRS-IN2P3, Universit\'e Joseph Fourier and INPG, 38026 Grenoble, France}\\
(3) {\it Physics Department, Temple University}\\
{\it Barton Hall, 1900 N. 13th Street, Philadelphia, PA 19122 -6082, USA}\\
$\dag$ {\it E-mail: x.artru@ipnl.in2p3.fr}
}
\end{center}
\vskip 8mm

\begin{abstract}
Positivity restrains the allowed domains for pairs or triples of spin observables in polarised reactions.  
Various domain shapes in ${1\over2}+{1\over2}\to{1\over2}+{1\over2}$ reactions are displayed.
Some methods to determine these domains are mentioned and a new one based on the anticommutation between two observables is presented. 
\end{abstract}

\vskip 8mm
\setcounter{equation}{0}

%

\section{The spin observables} 

We consider the polarised $2\times2$ reaction
\be\label{reaction}
A+B \to C+D\,,
\ee
where $A$, $B$, $C$ and $D$ are spin one-half particles. An example is 
\be\label{reaction-ppLL}
\ppLL~.
\ee
The fully polarised differential cross section of (\ref{reaction}) can be expressed as
\be
{d\sigma \over d\Omega} = I_0 \ F\left( \Sv_A,\Sv_B,\check \Sv_C,\check \Sv_D \right)\,,
\label{basic:eq:dcs}
\ee
where $F$ contains the spin dependence. $\Sv_A$ and $\Sv_B$ are the polarisation vectors of the initial particles ($|\Sv| \le 1$). $\check \Sv_C$ and $\check \Sv_D$ are pure polarisations ($|\check\Sv|=1$)  \emph{accepted} by an ideal spin-filtering detector. They must be distinguished from the \emph{emitted} polarisations $\Sv_C$ and $\Sv_D$ of the final particles. The latter ones depend on the polarisations of the incoming particles, e.g.,
\be\label{outgoingPol}
\Sv_C = \nabla_{\check\Sv_C} F\left(\Sv_A,\Sv_B,\check\Sv_C,\check\Sv_D=0\right) \,/\,F\left(\Sv_A,\Sv_B,\check\Sv_C=0,\check\Sv_D=0\right) 
\ee
$F$ is given in terms of the \emph{Cartesian reaction parameters} \cite{Bourrely:1980mr} by 
\be
F\left( \Sv_A,\Sv_B,\check \Sv_C,\check \Sv_D \right)
= C_{\lambda\mu\nu\tau} \ S^\lambda_A \, S^\mu_B \, \check S^\nu_C \, \check S^\tau_D
\,.
\label{FCart}
\ee
In the right-hand side the $\Sv\,$'s are promoted to four-vectors with $S^{0} = 1$.
The indices $\lambda, \mu, \nu, \tau$, run from 0 to 3,
whereas latin indices $i$, $j$, $k$, $l$, take the values 1, 2, 3, or $x$, $y$, $z$.
A summation is understood over each repeated index.
$S^x, S^y, S^z$ are measured in a triad of unit vectors
$ \{ \xu, \yu, \zu \} $ which may differ from one particle to the other. A standard choice is to take $\zu$ along the particle momentum and $\yu$ common to all particles and normal to the scattering plane. 
For example,
$C_{0000} \equiv 1$,
$C_{xy00} \equiv A_{xy}$ is an initial double-spin asymmetry,
$C_{000y}$ is the spontaneous polarisation of particle $D$ along $\yu$,
$C_{0y0y} \equiv D_{yy}$ is a spin transmission coefficient from $B$ to $D$ and
$C_{00xy} \equiv C_{xy}$ is a final spin correlation.
\goodbreak\goodbreak\bigskip
\noindent
The Cartesian reaction parameters are given by
\be\label{basic:eq:corr-param}
C_{\lambda\mu\nu\tau} = \tr\{\,\mathcal{M}\,\left[\sigma_\lambda(A)\otimes\sigma_\mu(B)\right]\,\mathcal{M}^\dagger
\left[\sigma_\nu(C)\otimes\sigma_\tau(D)\right]\,\}\ /\ \tr\{\,\mathcal{M}\,\mathcal{M}^\dagger\,\}\,,
\ee
which will be symbolically abbreviated as a sort of expectation value:
\be
(\lambda\mu|\nu\tau)\equiv C_{\lambda\mu\nu\tau}=\langle\sigma_\lambda(A)\,\sigma_\mu(B)\,\sigma_\nu(C)\,\sigma_\tau(D)\rangle~,\label{basic:eq:corr-param'}
\ee
with $\sigma_0=\mathbbm{1}\equiv\pmatrix{1&0\cr0&1}$. 

\section{The positivity constraints}

The cross section (\ref{basic:eq:dcs}) is positive for arbitrary \emph{independent} polarisations of the external particles, that is to say
\be\label{class-posit}
F\left( \Sv_A,\Sv_B,\check \Sv_C,\check \Sv_D \right)\le 1
\quad{\rm for}\quad 
\Sv_A,\ \Sv_B, \ \check\Sv_C,\ \check\Sv_D \in \hbox{unit ball}\ |\Sv|\le1\,.
\ee
%
However there are positivity conditions which are stronger than (\ref{class-posit}). The full positivity condition can be obtained from the positivity of the \emph{cross section matrix} $\R$ defined by   
\be\label{basic:eq:CSdM}\eqalign{  
\langle c,d | \mathcal{M}| a,b \rangle\,\langle a',b' | \mathcal{M}^\dagger| c',d' \rangle 
&=\langle a',b';c',d'|\R|a,b\,;c,d\rangle\cr
&=\langle a',b';c,d |\Rpt|a,b\,;c',d'\rangle
\,}\ee
in terms of the helicity or transversity amplitudes $\langle c,d | \mathcal{M}| a,b \rangle$.
$\Rpt$ is the partial transpose $\R$, the transposition $\R\to \Rpt$ bearing on the final particles\footnote{%
Alternatively, keeping the same $\Rpt$, one may define $R$ as the full transpose of that given by (\ref{basic:eq:CSdM}). Then the partial transposition between $\Rpt$ and $R$ would bear on the initial particles. This choice was done in Ref.\cite{Artru:2004jx}, where $R$ is called ``grand density matrix''.
}. 
All spin observables of reaction (\ref{reaction}) can be encoded in $\R$ or $\Rpt$. 
The diagonal elements of $R$ or $\Rpt$ are the fully polarised cross sections when the particles are in the basic spin states. By construction, $\R$ (but not necessarily $\Rpt$) is \emph{semi-positive definite}, that is to say $\langle\Psi|\R|\Psi\rangle\ge0$ for any $\Psi$. 

Equations (\ref{basic:eq:dcs}), (\ref{FCart}) and (\ref{basic:eq:corr-param}) can be rewritten as:
%
\be\label{basic:eq:dpcsR}\eqalign{%
{d\sigma\over
d\Omega}\left(\rho_A,\rho_B,\acc_C,\acc_D \right) &=
\tr\{\Rpt\ [\rho_A\otimes\rho_B\otimes\acc_C\otimes\acc_D]\,\}~,\cr
\phantom{{1\over1}} C_{\lambda\mu\nu\tau} &= \tr\{\Rpt\left[
\sigma_\lambda(A)\otimes\sigma_\mu(B)\otimes\sigma_\nu(C)\otimes\sigma_\tau(D)\right]\}\, /\,\tr \Rpt~,\cr
\phantom{{1\over1} C_{\lambda\mu\nu\tau}} &= \tr\{\R\left[
\sigma_\lambda(A)\otimes\sigma_\mu(B)\otimes\sigma^t_\nu(C)\otimes\sigma^t_\tau(D)\right]\}\, /\,\tr \R~,}
\ee 
with $\rho={1\over2}(\une+\Sv\cdot\sigma)$, $\ \acc={1\over2}(\une+\check\Sv\cdot\sigma)$. 
The last two equations of (\ref{basic:eq:dpcsR}) can be inverted as 
\be\label{basic:eq:csdm}\eqalign{%
{\Rpt}_1\equiv(2^4/\tr\Rpt)\ \, \Rpt\ &=\ C_{\lambda\mu\nu\tau}\ \sigma_\lambda(A)\otimes\sigma_\mu(B)\otimes\sigma_\nu(C)\otimes\sigma_\tau(D)~,\cr
{\rm or}\quad {\R}_1\equiv
(2^4/\tr\R)\ \, \R\ &=\ C_{\lambda\mu\nu\tau}\ \sigma_\lambda(A)\otimes\sigma_\mu(B)\otimes\sigma^t_\nu(C)\otimes\sigma^t_\tau(D)~.}
\ee
The matrix ${\Rpt}_1$ is normalised to have the same trace as the unit matrix and is directly obtained from $F$ replacing the $S^\mu$'s by $\sigma^\mu$'s. 

\section{Various domains for pairs of observables}

For one observable, for example $\Ocal=C_{0\mu0\nu}\equiv\langle\sigma_\mu(B)\,\sigma_\nu(D)\rangle$ we have the trivial 
positivity condition $\Ocal\in[-1,+1]$. For a \emph{pair} $\{\Ocal_1,\Ocal_2\}$ of such observables we have therefore $\{\Ocal_1,\Ocal_2\}\in[-1,+1]^2$.
However, in many cases the allowed domain is more restricted than the square. An empirical but systematic method \cite{Richard:1996bb,Elchikh:1999ir} to find the domain simply consists of generating random, fictitious helicity or transversity amplitudes, computing the observables and plotting  the results the one against the other. 
Once the contours revealed, it is an algebraic exercise to demonstrate rigorously the corresponding inequalities. 
Table~\ref{excl:tab:reca} summarises the shapes of the domains for the sixteen independent observables of the reaction (\ref{reaction-ppLL}). These domains are either the full square $[-1,+1]^2$ or the unit disk or a triangle.  
\begin{table}[!ht]
\caption{\label{excl:tab:reca} Domain allowed for pairs of observables: the entire square
$\left(\protect\sq\right)$, the unit disk $ \left(\protect\dis\right)$, the triangle $|2\Ocal_1| \le\Ocal_2+1$ $\left(\protect\trv\right)$, or
$|2\Ocal_2| \le \Ocal_1+1$ $\left(\protect\trh\right)$, where $\Ocal_1$ is horizontal and $\Ocal_2$ vertical. The symbol \protect\disp\ indicates that the pair of observables is constrained in the unit disk, but the corresponding operators do not anticommute. }
\begin{center}
\scalebox{.9}{
\begin{tabular}{ c c c c c c c c c c c c c c c c c c| c}
\rotatebox{90}{$A_{n}$}&
\rotatebox{90}{ $C_{nn}$}&
\rotatebox{90}{$D_{nn}$}&
\rotatebox{90}{$K_{nn}$}&
\rotatebox{90}{$C_{ml}$}&
\rotatebox{90}{$D_{mm}$}&
\rotatebox{90}{$C_{mm}$}&
\rotatebox{90}{$C_{ll}$}&
\rotatebox{90}{$D_{ml}$}&
\rotatebox{90}{$K_{mm}$}&
\rotatebox{90}{$K_{ml}$}&
\rotatebox{90}{$C_{nlm}$}&
\rotatebox{90}{$C_{nml}$}&
\rotatebox{90}{$C_{nmm}$}&
\rotatebox{90}{$C_{mnl}$}&
\rotatebox{90}{$C_{mln}$}&
\rotatebox{90}{$C_{mnm}$}&
\rotatebox{90}{$C_{mmn}$}\\
 \hline
\sq 	 &\trh  &\sq   &\sq  & \dis  & \disp  &\dis &\dis & \disp &\dis  &\dis   &\dis  & \dis & \dis & \dis &\dis &\dis & \dis&$P_{n}$  \\
       & \trh  & \sq & \sq &  \sq & \dis   &\sq &\sq  & \dis  &\dis  &\dis   & \sq  &\sq   &\sq   &  \dis &\dis &\dis & \dis&$A_{n}$ \\
       &         &  \sq& \sq & \trv &   \dis &\sq   &\sq   &\dis & \dis &\dis  &\sq   & \sq  & \trv  & \dis &\dis &\dis & \dis&$C_{nn}$ \\
      &          &       &  \sq & \dis &\sq   & \dis  &\dis  & \sq  &\dis &  \dis &\dis  &  \dis  &\dis &\sq&\dis&\sq&\dis &$D_{nn}$ \\
     &          &        &         &\dis  & \dis &\dis  & \dis &\dis  &\sq  &\sq   & \dis  & \dis   &\dis &\dis&\sq&\dis&\sq&$K_{nn}$  \\
    &           &        &          &      &\dis   & \dis  & \dis &\disp &\disp&\dis  &\sq   & \sq  &    \dis   &  \dis &\dis &\dis & \dis&$C_{ml}$\\
    &          &        &           &      &      &  \sq&  \dis &  \dis     &  \sq&\disp   &\dis    & \sq &   \dis  & \dis &\dis &\dis & \dis&$D_{mm}$ \\
    &          &        &          &       &      &      &   \sq&\dis      &\sq&  \dis  &\dis   & \dis  & \sq   &\sq&\sq&\dis&\dis& $C_{mm}$ \\
    &         &        &          &       &       &      &       &\sq     &\dis &\sq &  \dis   & \dis  & \sq  &\dis&\dis&\sq&\sq&$C_{ll}$ \\
    &        &          &          &      &      &      &       &          &\disp&\sq  &\sq  &  \dis   &  \disp   & \dis &\dis &\dis & \dis&$D_{ml}$ \\
    &        &         &          &       &     &       &      &          &      &   \dis     & \sq    &  \dis   &  \dis  & \dis &\dis &\dis & \dis&$K_{mm}$ \\
    &   &   &   &   &   &   &   &   &   &           & \dis  &\sq   & \disp&  \dis &\dis &\dis & \dis&$K_{ml}$  \\
    &   &   &   &   &   &   &   &   &   &   &            &\sq   & \dis& \dis&\sq&\sq&\dis&$C_{nlm}$  \\
      &   &   &   &   &   &   &   &   &   &   &            &  & \dis& \sq&\dis&\dis&\sq&$C_{nml}$  \\     
    &   &   &   &   &   &   &   &   &   &   &            &  & &  \dis &\dis &\disp & \disp& $C_{nmm}$  \\
       &   &   &   &   &   &   &   &   &   &   &            &  & & &\sq &\dis&\disp & $C_{mnl}$\\
  &   &   &   &   &   &   &   &   &   &   &            &  & &     & & \disp &\dis & $C_{mln}$\\
     &   &   &   &   &   &   &   &   &   &   &            &  & & & & & \sq& $C_{mnm}$\\
\end{tabular}}
 \end{center}
\end{table}

\subsection{Anticommutation method}\label{basic:sub:antic}
Disk-shaped domains are, in many cases, straightforward results of anticommutation of the observables of the pair. From the last equation of (\ref{basic:eq:dpcsR}), one can consider the observables as expectation values of operators. Since each $ \sigma_\mu^2$ is equal to the identity, we have $\Ocal^2=\mathbbm{1}$. Furthermore two such operators $\Ocal$ and $\Ocal'$ differing by at least one index ($\lambda$, $\mu$, $\nu$ or $\tau$) either commute or anticommute. 

For pairs of anticommuting observables, disk domains result from the following theorem:
\be
\hbox{If } \Ocal^2=\Ocal'^2=\mathbbm{1}\hbox{ and }\Ocal\hbox{ and }\Ocal'\hbox{ are anticommuting, then }
\langle \Ocal\rangle^2+\langle \Ocal'\rangle^2\le1~.
\ee

\ni\textsl{Proof:} set $x=\sqrt{\langle \Ocal\rangle^2+\langle \Ocal'\rangle^2}$, $\langle \Ocal\rangle=ax$, $\langle \Ocal'\rangle=bx$. Then $a^2+b^2=1$ and $\langle a\Ocal+b\Ocal'\rangle=x$. From $\Ocal^2=\Ocal'^2=\mathbbm{1}$ and $\Ocal\Ocal'+\Ocal'\Ocal=0$ one gets $(a\Ocal+b\Ocal')^2=\une$ which means that $a\Ocal+b\Ocal'$ has eigenvalues $\pm1$. Its expectation value $x$ has to be within these eigenvalues, therefore
$x^2\le1$.

 Note that a disk can occur even if the observables commute, for instance if, due to some symmetry, $\Ocal_2$ has the same expectation value as another operator $\Ocal'_2$ which anticommutes with $\Ocal_1$ and $\Ocal_3$. Examples of this situation are indicated by crossed circles of Table 1.

\section{Various domains for triples of observables}

\begin{figure}[!ht]
\centerline{\includegraphics[width=.85\textwidth]{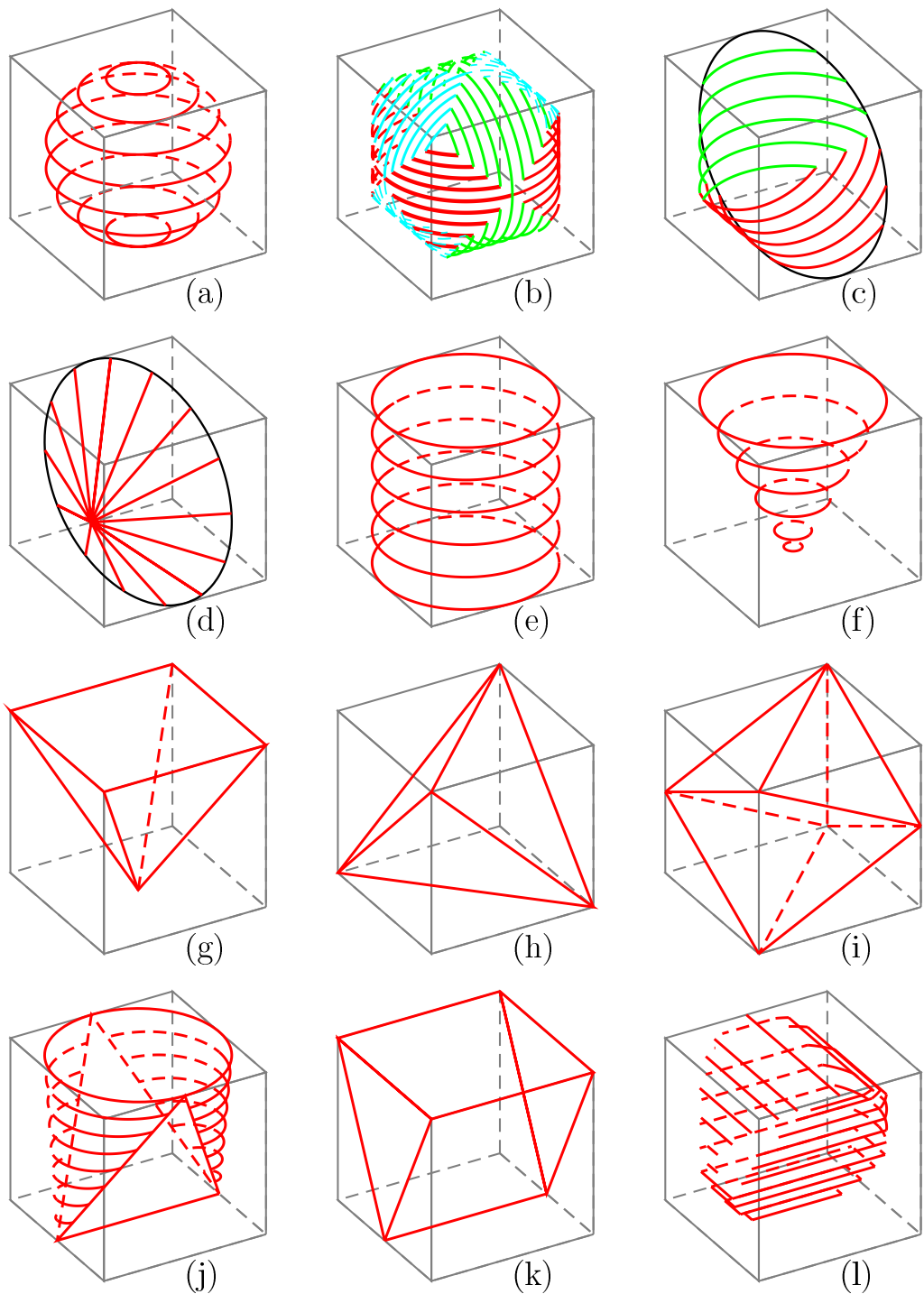}}
\caption{\label{excl:fig:lim3d}  Some allowed domains encountered in  simulating randomly three observables: 
the unit sphere (a), 
the intersection of three orthogonal cylinders of unit radius (b),
the intersection of two cylinders (c),
  or a slightly smaller  double cone (d), 
  a cylinder (e),
  a cone (f), 
a pyramid (g), 
  a tetrahedron (h),
  an octahedron (i),
  a ``coffee filter'' (j),
an inverted tent (k), 
and the intersection of two cylinders and a dihedral (l).  For clarity, part of the limiting surface is sometimes removed. Some figures transformed by parity with respect to the centre of the cube or by interchange of the axes are also obtained.}
\end{figure}

The empirical and anticommutation methods generalise straightforwardly to triple of observables. 
Figure~\ref{excl:fig:lim3d} shows the boundary of the domains that we have identified for the observables 
of the reaction (\ref{reaction-ppLL}):
the unit sphere, a pyramid, an upside-down tent, a cone, a cylinder, the intersection of two orthogonal cylinders or a double cone which is slightly smaller than this intersection,  a combination of the disk, square and triangle projections delimiting a volume similar to a ``coffee filter'', the intersection of three orthogonal cylinders (larger than the unit sphere!), a tetrahedron,  the intersection of two cylinders and two planes, an octahedron, or figures deduced by mirror symmetry.

\paragraph{Can the domain of a triple be the whole cube?}\label{basic:sub:limited}
Suppose now that for instance 3 observables $\Ocal_1$, $\Ocal_2$, and $\Ocal_3$, each of which has $+1$ and $-1$ as extreme eigenvalues, are commuting and that no symmetry relates a pair of them to a non-commuting pair. Does it means that their joint positivity domain $\D\{\Ocal_1,\Ocal_2,\Ocal_3\}$ is the whole cube? A partial negative answer is the following: If the reaction depend on $N$ independent amplitudes, $\D\{\Ocal_1,\Ocal_2,\Ocal_3\}$ can reach at most $N$ corners of the cube \cite{larevue}. The domains shown in Figure 1 are those of the reaction (\ref{reaction-ppLL}), which has $N=6$ and indeed none of them reaches more than 6 corners, this number being obtained for the domain (i). More generally, if $N<8$, all triple observables are restricted in domains smaller than the cube.

\section{Outlook}
\medskip

We have seen that the positivity restricts the pairs or triples of observables to subdomains of the square or the cube, some of which having non-trivial shapes. Here we have presented only two methods for determining these domains. Other methods use the Cauchy-Schwarz inequality or the positivity of the subdeterminants of $\R$ whose diagonal elements are on the diagonal of $\R$. For exclusive reactions, $\R$ is of rank one, therefore all diagonal $2\times2$ subdeterminants vanish. This links the observables by a large number of quadratic identities, from which inequalities can be obtained straightforwardly. We must tell, however, that inequalities expressing the positivity of $\R$ define joint domains for many observables and it is sometimes a straightforward but lengthy task to obtain the projected domain for two or three observables. 
 
We thank M. Elchikh and O.V. Teryaev for help, useful discussions and comments. 


%
\end{document}